# Atomic Conversion Reaction Mechanism of WO$_3$ in Secondary Ion Batteries

**Yang He[2], Meng Gu[1], Haiyan Xiao[3], Langli Luo[1], Fei Gao[4], Yingge Du[1*], Scott X. Mao[2,*] and Chongmin Wang[1*]**

[1]Environmental Molecular Science Laboratory, Pacific Northwest National Laboratory, Richland, Washington, 99352, USA

[2]Department of Mechanical Engineering and Materials Science, University of Pittsburgh, Pittsburgh, Pennsylvania, 15261, USA

[3]School of Physical Electronics, University of Electronic Science and Technology of China, Chengdu, 610054, China

[4]Department of Nuclear Engineering and Radiological Sciences, University of Michigan, Ann Arbor, Michigan, 48109, USA

[*]Corresponding authors: chongmin.wang@pnnl.gov (C.M.W.); yingge.du@pnnl.gov (Y.D.); sxm2@pitt.edu (S.X.M.)

Conversion reaction is one of the most important chemical processes in energy storage such as lithium ion batteries. While it is generally assumed that the conversion reaction is initiated by ion intercalation into the electrode material, solid evidence of intercalation and the subsequent transition mechanism to conversion remain elusive. Here, using well-defined $WO_3$ single crystalline thin films grown on Nb doped $SrTiO_3(001)$ as a model electrode, we elucidate the conversion reaction mechanisms during $Li^+$, $Na^+$ and $Ca^{2+}$ insertion into $WO_3$ by *in situ* transmission electron microscopy studies. Intercalation reactions are explicitly revealed for all ion insertions. With corroboration from first principle molecular simulations, it is found that, beyond intercalation, ion-oxygen bonding destabilize the W framework, which gradually collapses to pseudo-amorphous structure. In addition, we show the interfacial tensile strain imposed by the $SrTiO_3$ substrate can preserve the structure of an ultra-thin layer of $WO_3$, offering a possible engineering solution to improve the cyclability of electrode materials. This study provides a detailed atomistic picture on the conversion-type electrodes in secondary ion batteries.

Conversion-type lithium ion batteries (LIB) using electrodes such as transition metal oxides, hydrides, and sulfides are capable of utilizing all possible oxidation states of a compound, and thus can provide large specific capacity.[1] It is believed that the conversion reaction is initiated by ion intercalation process. For example, based on the morphology evolution, it is proposed that conversion mechanism in $FeF_2$ was a layer by layer reaction, which is initiated from the surface and propagates towards the inside.[2] The lithiation of $SnO_2$ was suggested to start with lithium ion insertion along a specific crystallographic direction, creating lithiation strips with dislocations.[3, 4] *In situ* observation of lithiation of $Co_3O_4$ reveals that lattice expansion during early stage of conversion reaction, which is attributed to lithium intercalation.[5] Though, no direct evidence of intercalation phase or valance state change of the transition metal was reported. In the case of lithiation of $RuO_2$[6], an intercalation phase were detected at the reaction front. X-ray and Raman spectroscopy revealed that Li at a low concentration can intercalate into monoclinic $WO_3$ for electrochromic application.[7, 8] These spatially ensemble average method obscure the relation between intercalation and conversion reaction. Thus, high spatial-resolved systematic microscopic study of microstructure and chemical state evolution from intercalation to conversion reaction is still lacking for the case of conversion-type electrode in secondary ion batteries. Also, it is intriguing whether the reaction mechanism holds when larger or multivalence ions, such as $Na^+$ and $Ca^{2+}$, are involved.

Tungsten trioxide ($WO_3$) is a widely studied electrochromic material as its pseudo-cubic cell contains a large, empty center site compared to that of a $SrTiO_3$ (Fig. 1a), providing an open environment for small ion (e.g., H, Li, and Na) reversible intercalation/removal. $WO_3$ has also recently received considerable interest as anode candidate for LIB and as a conversion-type electrode for next generation "large-ion" batteries, due to the spacious lattice channels and high valance state that can theoretically host maximum 6 alkali metal ions or 3 alkaline earth metal ions via fast conversion reaction.[9-11] Therefore, it is an ideal model system to study the intercalation and conversion reactions, and the interplay between these two chemical processes.

In this work, single crystalline $WO_3$ films grown epitaxially on Nb doped $SrTiO_3(001)$ (Nb-STO) substrates were used as testing electrodes. The heteroepitaxial structure provides high mechanical stability for high resolution TEM imaging, allows us to make electrical contact through conductive Nb-STO, and also enables us to study the impact of atomic defects and interfacial strain on the intercalation/conversion reactions. Solid cell setup (Fig. 1) was used for electrochemical ion ($Li^+$, $Na^+$, $Ca^{2+}$) insertion. With real-time atomic-scale imaging, nanobeam diffraction (NBD), electron energy loss spectroscopy (EELS) and first principle molecular simulation, an intercalation process is explicitly revealed for all ion species (Li, Na, and Ca). The intercalation process accompanies an insulator to metal transition, which consequently enhances the speed of subsequent conversion reactions. With extensive ion insertion, ion-oxygen bonding reduced transition metal framework, which gradually distorted and contracted into pseudo-amorphous W metal.

**Results and discussion**

General morphology evolution of $WO_3$ during $Li^+$ injection is shown by the sequential bright field TEM images in Figs. 1c-e. Electron diffraction indicates that as-grown $WO_3$ corresponds to monoclinic structure (space group P 21/c), which is in good agreement with our XRD results and other materials used in battery.[10] After the conversion reaction, the electron diffraction pattern (inset of Fig. 1e) indicates that the structure collapsed into amorphous W (a-W) metal and $Li_2O$. The reaction front (marked by the white dash lines) is roughly parallel to $WO_3$ [001] direction, implying that Li atoms diffused along [010] lattice channel. Different from many other conversion-type electrode materials,[4, 12] no obvious dislocation density increase, or "Medusa" zone, was observed at the propagating lithiation front, which can be attributed to the spacious lattice channels (3.8 Å) for ion conduction in $WO_3$. Assuming equal expansion along three <001> direction, the volume expansion at full lithiation was measured to be ~17%.

To reveal the reaction mechanism, phase and $W^{n+}$ valance state evolution across the reaction front were analyzed with high spatial-resolved NBD and EELS. Based on the scanning TEM (STEM) contrast and the appearance of diffused amorphous ring in the NBDs beyond the reaction front, the microstructure gradually lost crystallinity passing the reaction front, which literally should be the conversion front (CF). Referring to the NBDs and structures of $Li_xWO_3$[7], lithiated $WO_3$ symmetry gradually increases with increasing lithium insertion, featuring a monoclinic to cubic transition (space group evolution sequence: P21/c→Pm-3m) before CF (left side of CF in Fig. 2a). Spatial-resolved EELS spectrum (Figure 2c) shows that the $O_2$-edge of W shifted from 56.5 eV to 54.5 eV, indicating valance state decrease.[13] This spectroscopy information proves that Li intercalated into the $WO_3$ and reduced the $W^{6+}$. It is known that pure $WO_3$ is semiconductor with a bandgap of 2.8 eV and following lithium intercalation, $Li_xWO_3$ becomes metallic when $x>0.25$.[14] Since the electrochemical reaction needs injection of both electrons and ions, the intercalation induced semiconductor-to-metal transition can greatly enhance injection rate of Li ions, which consequently accelerate subsequent conversion reaction. Since $Li_xWO_3$ at low $x$ values (e.g. $x<0.36$) has similar NBD pattern with pristine $WO_3$ and weak Li $K$ edge signal in the EELS spectra, the depth of intercalation cannot be accurately measured. However, the distance between the CF and the point of cubic (Pm-3m, $Li_{0.93}WO_3$) transformation is ~ 60 nm, meaning that the intercalation depth should be larger than 60 nm. The spatial-expanded and catalyzing intercalation region before conversion reaction represents a similar "solid solution region" as intercalation-type $LiFePO_4$ cathode.[15, 16] This implies that "solid solution region" maybe common in spacious lattices as a result of kinetic diffusion of small lithium atoms.

In the conversion region, the NBD spots of $Li_xWO_3$ lattice gradually shifted outward relative to a-W ring (point 3 in Fig. 2b), meaning that the W framework contracted and finally collapsed (point 4 in Fig. 2a-b). Since the physical volume expansion with lithium insertion is relatively small, this solid state amorphization might be attributed to the chemical potential increase with reduction of $W^{n+}$ and Li-O bonds formation. To provide clear insights into the structural changes caused by lithium insertion, *ab* initio molecular dynamics simulations were performed on $WO_3$ with 2 Li atoms being inserted into the body-centered and face-centered sites of the unit cell. The systems were completely relaxed until it reaches a thermal equilibrium state. Figure 3a and b show the relaxed atomic arrangement and corresponding radial distribution functions (RDF) of $Li_2WO_3$. It is noted that <W-W> peaks appear around 2.7 Å, which is very close to the bonding distance of 2.74 Å in W metal, indicative of the formation of W metal as the lithium/tungsten ratio reaches 2. We also find that <Li-O> bonds are formed in $Li_2WO_3$, as indicated by the appearance of <Li-O> peaks around 2.0 Å, which is comparable to the experimental value of 1.996 Å in $Li_2O$. These results are shown to agree well with experimental observations.

As a result of the congruent contraction of local W-framework, ultrafine sized W crystals were formed. The HRTEM images of the fully lithiated phase (Fig. 4a-b) show no crystal feature but "clusters" of 2~5 Å in size. The diffused diffraction pattern (Fig. 4c) is consistent with the simulated diffraction of polycrystalline W metal with 3 Å grain size. The unit cell of W metal is ~3.16 Å. This result means that the W atoms have short range order of pure W metal, i.e. pseudo-amorphous structure. On the other hand, radiating with high intensity (300 KeV, $\phi \sim 6 \times 10^{19}\,e \cdot cm^{-2}s^{-1}$ ) electron beam, the pseudo-amorphous structure was gradually transformed into W nanocrystals (Fig. 4d-e). The diffraction pattern of the radiated region (Fig. 4f) shows sharp rings, which match with the simulated diffraction of polycrystalline W metal (average grain size 3 nm). This crystallization may be attributed to electron beam induced heating and displacement effects.[17] Based on *in situ* TEM studies, it has been commonly noticed that for conversion electrodes, nanocrystalline transition metals were formed after charging,[5, 18] contradicting to the pseudo-amorphous structures observed by other techniques.[1, 6, 19] Based on our present analysis, we notinced that the converted region should be pseudo-amorphous, but high dose  electron beam exposure appears to trig the amorphous to crystalline transformation.

Reviewing above findings, the whole picture of the conversion reaction upon $Li^+$ injection into $WO_3$ electrodes can be summarized as follows. The $Li^+$ ions diffused along the vacant channel of the lattice, intercalated into the $WO_3$ unit cells and reduced $W^{6+}$. The lattice gradually evolves from monoclinic to cubic symmetry and is accompanied by increasing in conductivity, which effectively facilitates subsequent conversion reaction as Li concentration increases. Beyond intercalation, $Li^+$ ions bonded with oxygen atoms, further reduced $W^{n+}$ and destabilized the structure. W framework shrunk towards W metal clusters. At full charge, the $W^{n+}$ ions were reduced to $W^0$ and the W framework collapsed into pseudo-amorphous W metal, Li-O bonds were established to form $Li_2O$ structure.

To test whether the above mechanism holds for larger and/or multivalence ions, $Na^+$ (radii 95 pm vs. $Li^+$ radii 60 pm)[20] and $Ca^{2+}$ (radii 109 pm) insertions into $WO_3$ were studied. Figure 5 shows sequential atomic-resolution images of $Na^+$ insertion in $WO_3$. Similar to the case of Li ion, the CF followed {001} planes, indicating sodium ion diffuse along <001> channels. The CF progressively propagated upward from the bottom contact with $Na^+$ source. Fast Fourier transformations (FFT) of the HRTEM images were used to analyze the phases across the CF. The FFT of the region far away from the CF (Fig. 5d) can be assigned to both $WO_3$ and $Na_xWO_3$ phases (x ~ 0.1-0.74, oxygen octahedral tilted). The FFT of the region right above the CF (Fig. 5e) can only be assigned to $Na_xWO_3$ (x ~ 0.11-1, Pm-3m space group, oxygen octahedral aligned). Since $Na_xWO_3$ conductivity also increases with increasing $x$ values,[21] the Na intercalation should also facilitate subsequent conversion reaction. These results support an intercalation-induced phase evolution similar to the case of Li. The HRTEM image and FFT of the converted region (Fig. 5c, f) shows W nanocrystals embedded in amorphous $Na_2O$. Since pseudo-amorphous structure was identified in areas with minimal beam exposure, the formation of large W metal crystals in Figure 5c should be attributed to the imaging electron induced effect. Above all, $Na^+$ insertion in $WO_3$ shows similar intercalation-conversion reaction mechanism as that for $Li^+$.

$Ca^{2+}$ injection in $WO_3$ was also proved to follow the interaction-conversion reaction as discovered for the case of Li and Na as described above. Figure 6c-e show HRTEM, HRSTEM and EELS of the region above CF. Comparing with pristine lattice (Fig. 6b), the enhanced contrast at the centers of $WO_3$ lattices (Fig. 6c) proves $Ca^{2+}$ intercalation. Besides, EELS spectrum of this region shows sharp Ca $L$ edges (Fig. 6e). Unfortunately, the conversion only proceeded for ~2 nm into the film before fully stopped, which may be attributed to the poor conduction of $Ca^{2+}$ through the denser W metal framework (Fig. 6d).

Another observation worth mentioning is that one monolayer (ML) of $WO_3$ was preserved at the heteroepitaxial interface after the conversion reaction was finished. For the case of lithiation, the two W planes in the preserved layer shifted by half unit cell as shown in Fig. 7b. Referring to the structure of interface, $WO_6$ octahedral shares one O atom with the $TiO_6$ octahedral, i.e. the W-O plane on interface is chemically constrained by the substrate. So, it is likely that the mechanical/chemical coupling at the strained interface preserved the structure integrity of $WO_3$ during lithium insertion. Note that it is still an open question if this ultrathin layer worked to accommodate Li atoms. Since lithiation induced structure collapse destroys lattice channels and bars practical application of many materials in lithium ion batteries, e.g. $CuF_2$,[18] this result points to potential engineering solutions to tune the chemical reactivity in batteries. Further investigation is underway.

**Conclusion**

Microscopic detail of conversion reaction in $WO_3$ during $Li^+$, $Na^+$ and $Ca^{2+}$ ions insertion were systematically studied. In all cases, the conversion reaction was initiated by intercalation. The intercalation reaction gradually increases structural symmetry and conductivity, facilitating subsequent conversion reaction. Beyond intercalation, $Li^+$ ions bonded with O, reducing $W^{n+}$ and destabilizing W framework which gradually shrunk to pseudo-amorphous W metal. Interestingly, the interface mechanical/chemical constrain "held" the integrity of the original W framework upon ion insertion. With unprecedented resolution, this work presents a clear systematic picture of conversion reaction in conversion-type electrode materials upon ion injection.

**Materials**

(001) oriented $WO_3$ thin films were grown on conductive Nb doped (0.7 wt%) $SrTiO_3$ (001) (Nb-STO) substrates by molecular beam epitaxy. Details on film growth and characterization have been described elsewhere.[22] Films with thickness above 200 nm were used in this study which were shown to exhibit a monoclinic structure as revealed by nanobeam diffraction, reflection high-energy electron diffraction, and high resolution X-ray diffraction.[22] The epitaxial interface guarantees the sample mechanical stability and ensures excellent contact with atomic level registry between $WO_3$ thin film (electrode) and Nb-STO (conductive binder).

**Methods**

The TEM sample was made as follows. Together with Nb-STO, the $WO_3$ film was cut by diamond saw, mechanical polished and Ar-ion milled at 5-2 KeV for thin cross-section. Then the cross-section sample was pasted onto Mo ring and loaded on sample side of the Nanofactory-STM holder. Pure metals (Li, Na and Ca) were used as ion sources which were loaded onto a tungsten tip and navigated by piezo-system in the Nanofactory holder. The metal loading process was carried out in an Ar filled glove box and the whole setup was transferred to TEM via a home-made vacuum container. The total air-exposure time was controlled to be <5 seconds. A thin layer of oxides formed on the metal sources surface working as solid electrolytes. FEI Titan 80-300 S/TEM with probe $C_s$ corrector and ETEM with imaging $C_s$ corrector (both operating at 300 kV) were used for this research. The sample was adjusted to [010] zone axis before experiment. Ion source was navigated to contact the $WO_3$ film then a bias (~-0.8V) was applied to drive the electrochemical reaction. The NBDs and EELSs were taken right after reaction without taking the sample out of TEM. The NBDs were captured at 300 kV microprobe scan mode with 50 μm condenser apertures (C2). The probe diameter was measured to be ~3 nm (full width at half maximum) with 0.9 mrad convergence angle. The *ab initio* molecular dynamics (MD) simulations were carried out using SIESTA code,[23] based on density-functional theory with the generalized gradient approximation and the Perdew–Burke–Ernzerhof functional.[24] The valence wavefunctions were expanded in a basis set of localized atomic orbitals and single-$\zeta$ basis sets were used. The simulations were conducted with a supercell containing 64 W and 192 O atoms, and 128 lithium atoms were initially inserted into the body-centered and face-centered sites of the $WO_3$ unit.

**Acknowledgement**

The in situ microscopic study described in this paper is supported by the Laboratory Directed Research and Development Program as part of the Chemical Imaging Initiative at Pacific Northwest National Laboratory (PNNL). YD acknowledges support by

**Figure Captions**

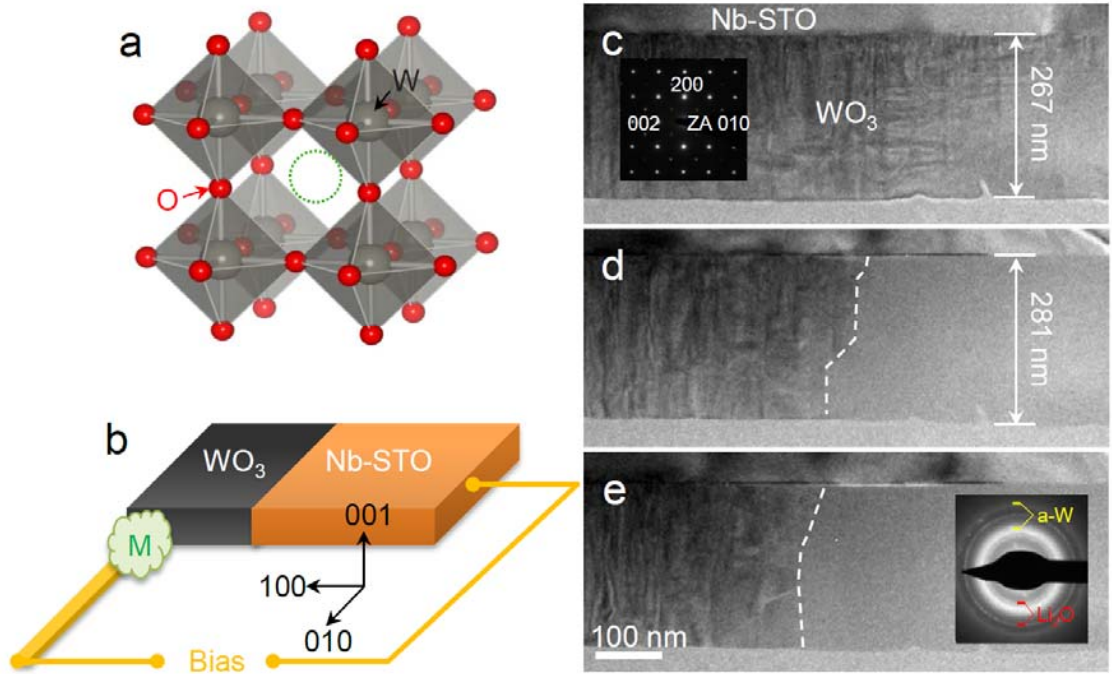

Figure 1, (a) atom structure of $WO_3$ showing $WO_6$ octahedral and vacant site, (b) experiment setup schematics, (M=Li, Na) (c-e) sequential TEM bright field images showing $WO_3$ structure evolution during lithiation. Insets show local electron diffraction patterns. White dash lines mark reaction front.

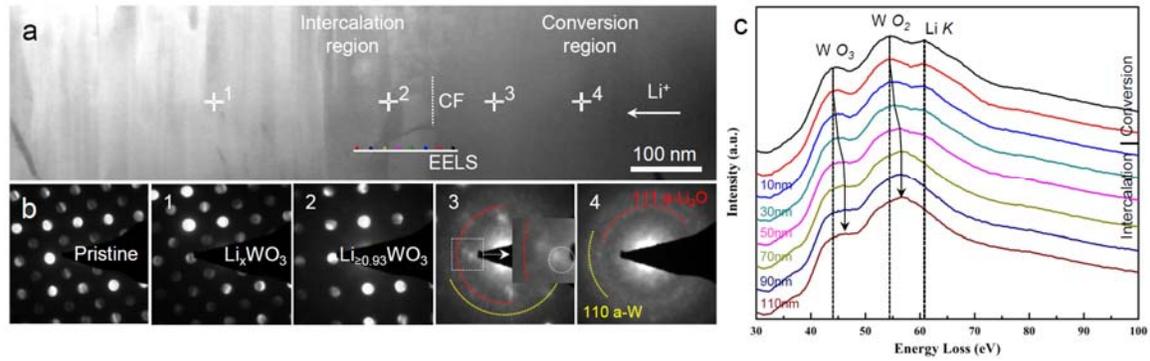

Figure 2, (a) STEM image of the CF in Figure 1e, (b) NBDs from the pristine sample and across the CF, (c) EELSs across the CF. Black arrow indicates the W $O_2$ peak shift. Note that the EELS spectrums have been aligned using zero-loss peak.

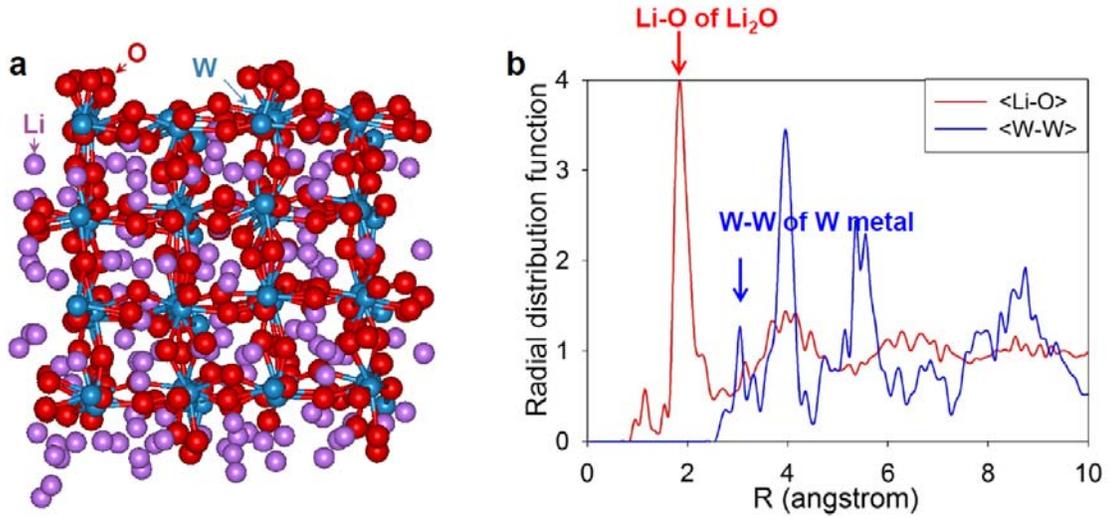

Figure 3, (a) schematic view of the relaxed Li$_2$WO$_3$, (b) RDFs for Li-O, W-O and W-W in the relaxed Li$_2$WO$_3$. Star markers indicate RDF positions of cubic Li$_x$WO$_3$ phase in which Li atoms take center position as shown in Fig. 1a. RDF positions of Li$_2$O and W metal were also marked.

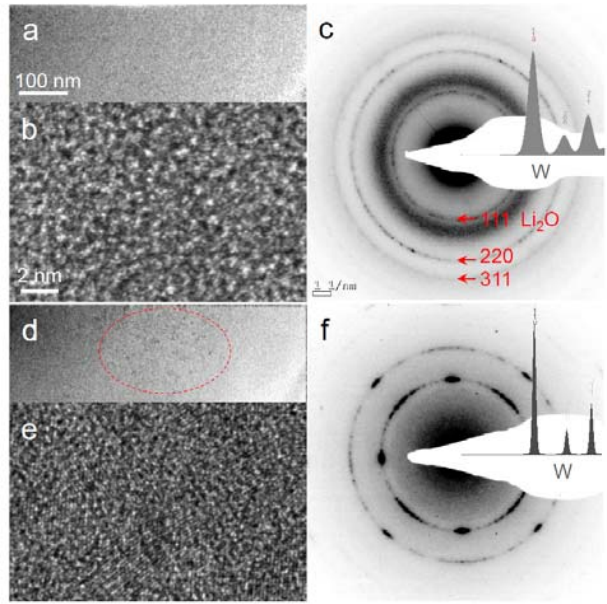

Figure 4, TEM, HRTEM and electron diffraction pattern of fully lithiated phase (b-c) and beam annealed phase (d-f). Insets in (c) and (f) are simulated diffraction intensity profiles.

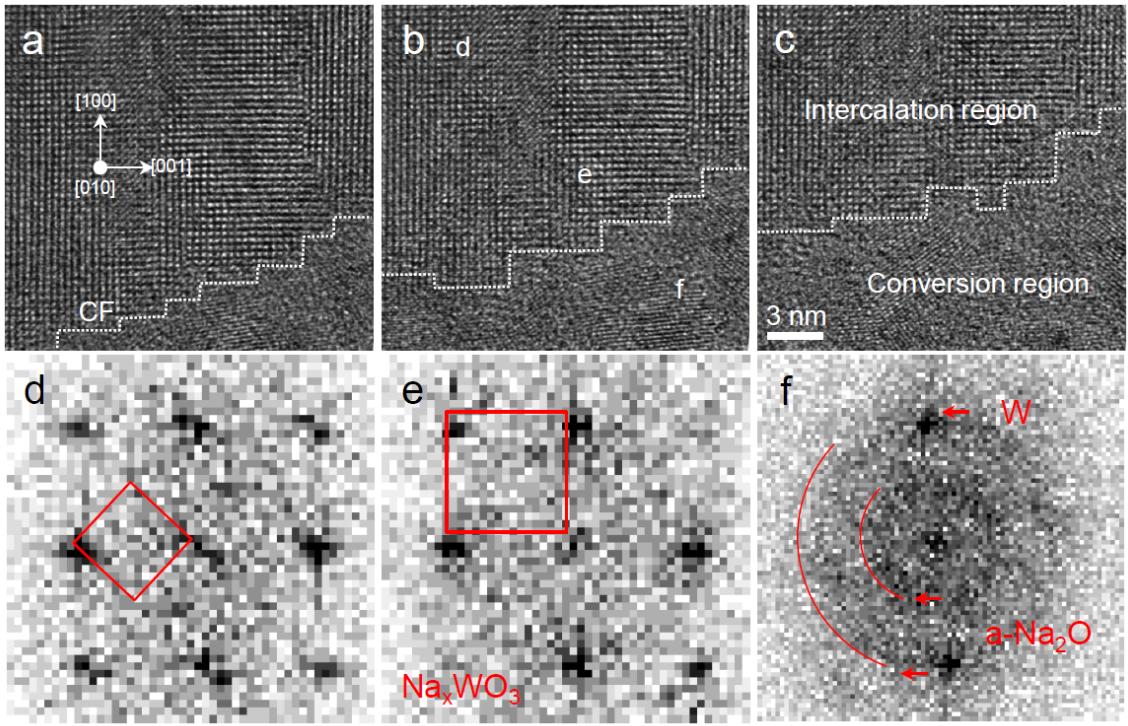

Figure 5, (a-c) sequential HRTEM images of conversion reaction during Na⁺ insertion in WO₃. Dash lines mark the CF. (d-f) FFTs across the CF in (b)

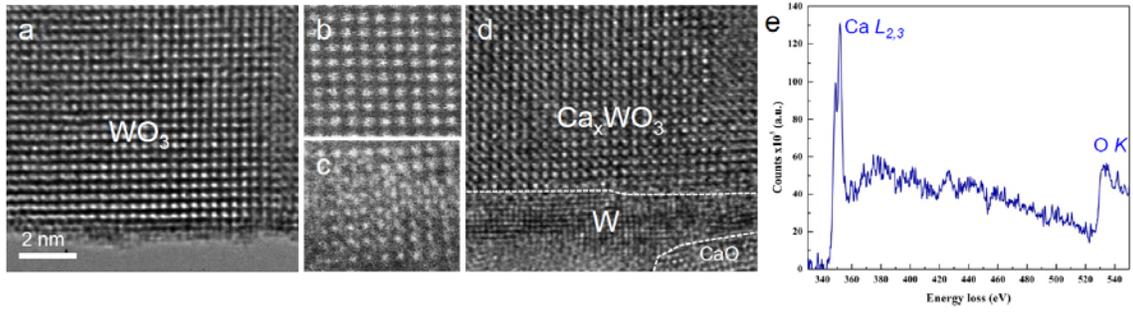

Figure 6, Atomic resolution HRTEM and STEM-HADDF images of WO3 before (a, d) and after (c, d) $Ca^{2+}$ insertion. (e) EELS of the intercalation region in (d).

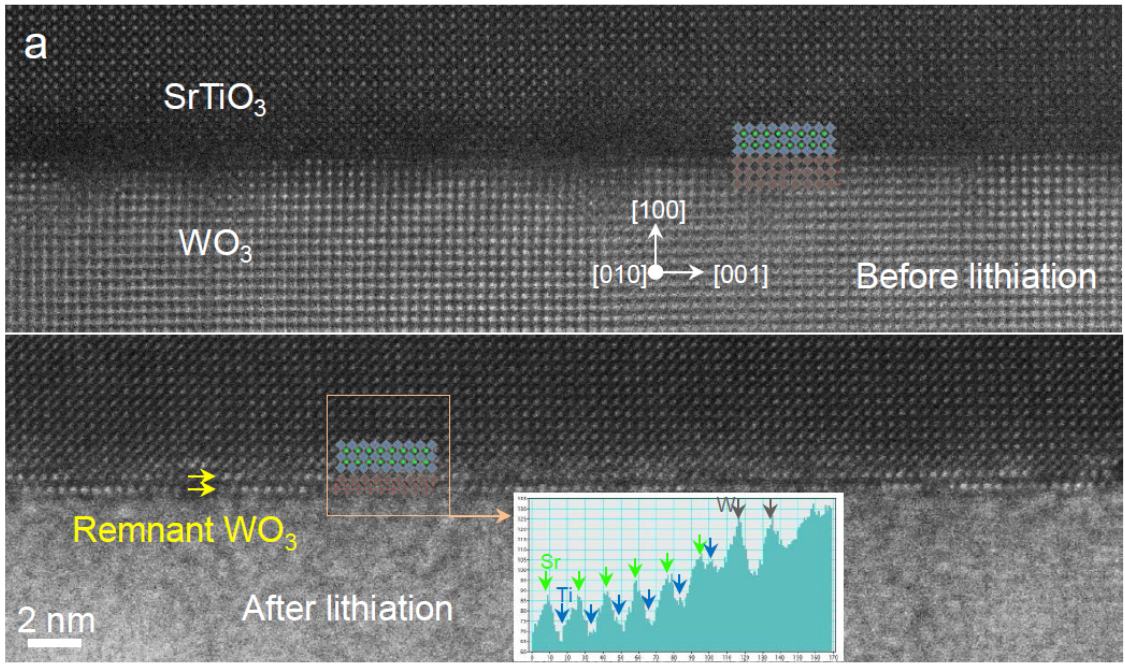

Figure 7, (a) Atomic resolution STEM-HADDF images of the SrTiO$_3$/WO$_3$ interface before (a) and after (b) lithiation. Insets show structure and extracted intensity profile across the interface.